\documentclass[preprint,showpacs,preprintnumbers,prl]{revtex4}
\begin{document}
\title {Nanoconfined Polystyrene: A New Phase}
\author {Sudeshna Chattopadhyay, Alokmay Datta, Avijit Das}
\affiliation {Surface Physics Division, Saha Institute of Nuclear Physics,
1/AF, Bidhannagar, Calcutta 700 064. India.}
\author {Angelo Giglia, Stefano Nannarone, Nicola Mahne}
\affiliation {TASC-INFM, AREA Science Park, ss 14 Km 163.5, I-34012,
Basovizza (TS), Italy}

\begin{abstract}
Transverse layering of molecular gyration spheres in spin-coated
atactic polystyrene ($aPS$) films, for film thickness $R \leq$
4$R_g$ ($R_g$ = unperturbed gyration radius), causes an increase in
free energy that does not follow the $(R_g/R)^{-2}$ dependence of
planar confinement and is explained by invoking a fixed-range,
repulsive, modified P\"{o}schl-Teller intermolecular potential, its
strength decreasing with increase in $R$. Vacuum ultraviolet
spectroscopy reveals a change in 'physical dimers' of adjacent
pendant benzene rings of $aPS$ from 'oblique' to 'head-to-tail'
configuration as film thickness goes from 9$R_{g}$ to 2$R_{g}$. This
reduces cohesion by reducing dimer dipole moment. Thus a new phase
of $aPS$, the nanoconfined phase, ordered but with lower cohesion
than bulk, is formed.

\end{abstract}
\maketitle

One-dimensional geometrical confinement of fluids causes the fluid to
form layers normal to the confinement direction \cite{layer,epl96,epl02a}.
For such 'nanoconfined' simple fluids the layer periodicity is equal to the molecular
size while for polymers (complex fluids) it is equal to the unperturbed gyration radius ($R_{g}$) \cite{epl96}, i.e., the dimension of a polymer molecule in the maximum entropy configuration \cite{DoiEdBk}. Nanoconfined fluids exhibit radically new
mechanical, thermal, dielectric and rheological  properties \cite{props,rev} . In a very recent study, a nanoconfined simple liquid
has been observed to be in a 'laterally cooperative'
state that behaves liquid-like or solid-like depending on the kinematics of the measurement process \cite {lan06}. In polymers layers form only when film
thickness is less than 4$R_{g}$ and there is a reduction in cohesion
between adjacent molecular gyration spheres \cite{prb05}, i.e. an
increase in free energy. These new properties suggest a basic reorganization at molecular levels and they are bound to have strong implications in any technology employing thin fluid films such as optoelectronic and magnetic coatings, adhesives, biological membranes and emerging nanotechnologies, in particular, photonics and nanofluidics.

In this communication we show that the increase in free energy due
to layering does not tally with that observed for planar confinement
\cite{Muthu,nl06}. We find, from tapping-mode atomic force
microscopy studies, a very similar confinement-induced reduction in
cohesion between adjacent gyration spheres on the film surface. We
have explained this drop in cohesive interaction as the emergence of
a new, repulsive intermolecular potential that fits well with a
modified P\"{o}schl-Teller potential(MPT) \cite{FlBk}, whose depth
can be increased by thinning the film but whose range is invariant
with confinement. We have also shown, through vacuum ultraviolet
(vuv) spectroscopy, that confinement causes a change in the geometry
of pairs ('physical dimers') of adjacent pendant benzene rings in
polystyrene from 'oblique' to 'head-to-tail' that reduces the dipole
moment of each 'dimer', which in turn, reduces cohesion between
molecular gyration spheres.

Atactic polystyrene (\emph{aPS}, mol.~wt.~$M\simeq 560900$,
$R_{g}=0.272M^{\frac{1}{2}}\simeq$ 20.4~nm) \cite{IsraBk} was
spin-coated on fused quartz plates from toluene solutions using a
photo resist spin-coater (Headway Inc.) to form films with thickness
($R$) varying from 40~nm ($\simeq 2R_{g}$) to 180~nm ($\simeq
9R_{g}$), and with air/film and film/substrate interfacial roughness
$\sim$ 0.6~nm, as has been described previously \cite{prb05}.
Contaminants are removed from the substrate by boiling it with 5:1:1 $H_2 O$ : $H_2 O_2$ : $NH_4 OH$ solution for 10 minutes, followed by rinsing in acetone and ethyl alcohol.

Atomic Force Microscopy (AFM) images were acquired in tapping-mode
with Nanoscope IV, Veeco Instruments, using etched Si tip and
Phosphorus doped Si cantilevers. The free amplitude was
$\textsf{A}_0$ = 36~nm, set point amplitude $\textsf{A}$ = 10.94~nm,
cantilever quality factor $Q$ = 505, resonance frequency $\omega_0$
= 2$\pi$~283~kHz and spring constant $k$ = 20~Nm$^{-1}$. x-ray reflectivity (XR) data of polystyrene films were
collected using the Cu $K_{\alpha1}$ line ($\lambda$ = 0.1540562~nm)
from an 18~kW rotating anode x-ray generator (Enraf Nonius FR591),
and Electron Density Profiles (EDPs) along film thickness were
obtained using standard methods of analysis \cite{epl96,prb05}.

Figure 1(a) shows the reflectivity profiles (open circles) of
\emph{aPS} films of different $R$-values, and the extracted EDPs from
best fits (line) are shown in Figure 1(b) in the same sequence and
having the same color code. For $R \leq$  4$R_g$ (84.0~nm) we
observe formation of layers in \emph{aPS} parallel to the substrate
surface, the error in $\rho$ being an order of magnitude less than
this variation \cite {epl96}. The reduction in cohesive energy caused
by the variation of density due to layering is given by \cite {IsraBk, Hoff, prb05}
\begin{eqnarray}
\Delta G^{(o)}_{PS-PS} = -\Delta A_{PS}^{(o)} /(2.1\times10^{-21})
\nonumber
\\ = -\sigma_{PS} (\rho(z)^2 - \rho_{max}^2)/(2.1\times10^{-21})
\end{eqnarray}
where $\Delta G^{(o)}_{PS-PS}$ is the reduction in (out-of-plane)
cohesive energy caused by the variation of density due to layering,
$\Delta A_{PS}^{(o)}$ is the (out-of-plane) change in polystyrene
Hamaker constant, $\sigma_{PS}$ is the polarizability of \emph{aPS}
and $\rho(z)$ ($\rho_{max}$) denotes the electron density at depth
$z$ (corresponding maximum).

Figure 1(c) shows the variation of $\Delta A_{PS}^{(o)}$ with
$(R_g/R)$, obtained from Eqn (1). The continuous line is the best
fit to the data (open circles) with the function
\begin{equation}
\Delta A_{PS}^{(o)} = K(R_g/R)^b
\end{equation}
The value of $b$ obtained from this fit is 3.0~$\pm$~0.3. This
deviates clearly from $b =$ 2, for an ideal polymer, or from $b =$
1.7, for a self-avoiding polymer \cite{Muthu}, under planar
confinement \cite{nl06} and, rather, correspond to spherical
confinement \cite{Muthu} of the polymer. We are thus led to study
the variation of surface free energy of the polymer films with $R$
to inspect for any deviations from planar confinement.

Figures 2(a) and 2(b) show the phase images obtained from
tapping-mode AFM of two typical \emph{aPS} films with $R \simeq$
50.0~nm and 84.0~nm, respectively. The topographical images of all
these films show roughly spherical features with an average diameter
of $R_{g}$ \cite{prb05}, corresponding to gyration spheres. The
frusta ($\simeq$ 0.6 nm high) of these spheres are consistent with
the top roughness obtained from x-ray studies. But the phase images
show larger variations in phase-shifts between adjacent 'spheres' on
film surface as $R$ reduces from 84.0~nm to 40.0~nm, implying a
larger variation in energy dissipated by the AFM tip in going over
from sphere to sphere \cite{Tam} and, by extension, a spatial
variation in surface free energy that increases with decrease in
$R$. This spatial variation is not observed for $R > 4R_{g}$ and it
cannot be explained by simple planar confinement of the polymer.
Thus, above a certain degree of confinement, the very nature of
confinement is changed by the formation of layers. We have tried to
find out what exactly is changing in the films from the layering
induced variations in free energy along film-depth, presented above,
and on film-surface, discussed below.

To the end of determining the variation in surface free energy
caused by layering, we have estimated the average energy dissipated
per cycle by the tip over the film surfaces, $E_D$, using the
expression \cite{Tam}

\begin{equation}
\sin\phi=(\frac{\omega}{\omega_{0}}\frac{\textsf{A}}{\textsf{A}_{0}})+\frac{QE_{D}}{\pi
k \textsf{A} \textsf{A}_{0}}
\end{equation}

where $\phi$ is the phase-shift with respect to the drive signal and
the other terms have been described above.

The tip exerts a van der Waals force on the surface, during approach
and retraction. This interaction is modeled as a sphere approaching
a plane with an effective contact area 4$\pi r_c \alpha_{Si}$, where
$r_{c}$ is radius of tip-curvature ($\simeq$ 10nm) and $\alpha_{Si}$
is the Si atomic diameter (= 0.22nm). Then energy dissipation by the
tip in the film planes with respect to minima is given by
\cite{IsraBk, prb05a}
\begin{equation}
\Delta E_{D}= \frac{2}{3}\frac{r_{c} \alpha_{Si}}{z_{0}^{2}} \Delta
A_{SiPS}
\end{equation}
where tip-sample adhesion, expressed by $A_{SiPS}$ (the
corresponding Hamaker constant) is considered to be the varying
interaction \cite{prb05a}, which is unaffected by cantilever tilt
\cite {Langmuir04}. Here $z_0$ is the tip-sample separation
($\simeq$ 0.2 nm in contact \cite{IsraBk}). Using the expression
$A_{SiPS} = A_{Si}^{\frac{1}{2}}A_{PS}^{\frac{1}{2}}$
\cite{prl02epj03}, where $A_{Si}$ and $A_{PS}$ denote the Hamaker
constants of Si and \emph{aPS} respectively, along with Eqns (3) and
(4), the value of $A_{Si}$ \cite{prl02epj03} and some simple
algebra, we determine $\Delta A_{PS}^{(i)}$, the in-plane variation
in \emph{aPS} Hamaker constant and hence $\Delta G_{PS-PS}^{(i)}$,
the (in-plane) variation in cohesion.

Figure 1(d) shows the observed variation (in solid circles) of
$\Delta G^{(o)}_{PS-PS}$ with depth $z$ across the gap between
adjacent layers, for all the different $R$-values probed. Similarly,
Figure 2(c) depicts $\Delta G^{(i)}_{PS-PS}$ variation (solid
circles) over adjacent gyration spheres as a function of in-plane
coordinate $x$, for different film thicknesses. In both cases the
abscissae have been shifted arbitrarily for clarity. It is
interesting to note that in all cases $\Delta G_{PS-PS}$'s have a
form that fits very well with the MPT potential given by \cite{FlBk}
\begin{eqnarray}
\Delta G_{PS-PS}(\xi) = V_0 \cosh^{-2}\alpha\xi = g^2 \alpha^2
\cosh^{-2}\alpha\xi
\end{eqnarray}

Here $V_0$ is the peak strength of the repulsive intermolecular
potential, which has a quadratic dependence on $\alpha =
\Lambda^{-1}$, $\Lambda$ being the range of the potential (at $\xi
=2\Lambda$, $V = 0.07 V_{0}$ ), and $g^2$ has the dimension of
energy. The best fit curves of data with the MPT potential are shown
in continuous lines in Figures 1(d) and 2(c) and values of $V_{0}$
and $\Lambda$ obtained from these fits are given in Table 1. From
this table it is clear that confinement has introduced an additional
intermolecular potential whose magnitude, given by $V_{0}$,
increases as film thickness is decreased but whose range remains
more-or-less invariant. It should also be noted that, for a film
with thickness $>$ 4$R_g$ the in-plane potential is measurable but
very small, consistent with the out-of-plane measurements. The
situation is depicted in the cartoon in Figure 2(d).

In order to correlate this new intermolecular potential with some
specific change in the molecular configuration of \emph{aPS}, we have
carried out vuv spectroscopy of polystyrene films with $R \simeq$
2$R_g$ and $R \simeq$ 9$R_g$. Transmission spectra in the 4-9~eV
range were collected in 10~meV steps at BEAR beamline of ELETTRA
synchrotron, with nearly linearly polarized light (the estimated
Stokes parameter $S_1 \simeq$ 0.5), the electric field lying in the
film plane \cite{bear}. The experimental chamber was maintained at
$\sim$ 10$^{-10}$~Torr and all measurements were done at ambient
temperature. Our focus was on the pure electronic singlet transition
$^1A_{1g} \rightarrow ^1E_{1u}$ involving the pendant benzene rings
of \emph{aPS}, which is centered around 6~eV.

Figures 3 show this spectral band for 180.0~nm (a) and 50.0~nm (b)
thick \emph{aPS} films. The split in the band can be explained as
arising from the resonant transfer interaction between correlated
clusters of pendant benzene rings, given by $J_{\beta} = \Delta\nu/2
\simeq 428~meV$, where $\Delta\nu$ is the measured split
\cite{PopeBk}, which causes the mixing of the singly excited states
of individual benzene rings through their transition dipole moments.
The doublet splitting indicates that 'dimers' of benzene rings are
involved in these clusters. The intensity ratio of the high energy
(-) and low energy (+) components of the doublets, $I_+ / I_- = (1 +
\cos\alpha)/(1 - \cos\alpha) = \cot^2(\alpha/2)$ gives $\alpha$, the
angle between the transition dipoles, i.e. the dihedral angle
between rings of the 'dimer' \cite{cpl04}, since the transition
dipole is entirely in the ring-plane. The strong electronic band
gives a clear indication of the change in the value of $\alpha$
between the two films from the change in intensities of these
components. $\alpha$ goes from $\simeq 75^{\circ}$ to $\simeq
0^{\circ}$ as film thickness goes from 180.0~nm to 50.0~nm,
corresponding to an 'oblique' or \emph{ob} configuration (shown
schematically in inset, Figure 3(a)) and a 'head-to-tail' or
\emph{ht} configuration (inset, Figure 3(b)) \cite{PopeBk},
respectively. A benzene 'dimer' has a permanent dipole moment only
when rings are non-parallel \cite{jcp75}, hence the \emph{ht}
'dimer' will have near-vanishing dipole moment. This would make it
undetectable through standard spectroscopic techniques \cite{jcp75}
and to our knowledge this is the first direct experimental evidence
of this 'near-parallel' benzene 'dimer'. Reduction in 'dimer' dipole
moment due to this configurational change would reduce coupling
between gyration spheres containing such 'dimers'. We suggest that
this is manifested as the repulsive MPT intermolecular potential.

We have found a completely new phase of atactic polystyrene under confinement - the \emph{nanoconfined phase}, more ordered than the (inherently disordered) bulk but less cohesive. Observation of similar phases in a simple fluid \cite{epl02a} indicates the universality of this phase and also shows a limit to which simple and complex fluids have the same behavior. We show here that in \emph{aPS} this phase is achieved through the alignment of adjacent benzene rings, explaining the similarity of confined \emph{aPS} to the helically ordered phases of syndiotactic \emph{PS}, observed in infrared spectra \cite{macro06}. The contradictory properties of this phase may explain its other observed properties, in particular its solid-liquid duality \cite{lan06}, reduction of $T_g$ with confinement and its dependence on $R_g$, and return to bulk $T_g$-value on adding small-molecule diluents \cite{Tg,rev}. This phase would also usher in new concepts in miscibility and solvation.

\begin{figure}
{\bf Figure captions} \caption{(color online) (a): X-ray
reflectivity data (circles, Fresnel reflectivity normalized and
upshifted) and fits (lines) of polystyrene (\emph{PS}) films on
quartz with different thicknesses $R$ (shown beside each curve).
(b): Electron Density Profiles (EDPs) along film depth from
reflectivity fits, color-coded and presented in same sequence. (c):
$\Delta A_{PS}^{(o)}$, increase in free energy due to layering,
versus $(R_g/R)$, $R_{g} =$ unperturbed gyration radius of \emph{PS}
(20.4~nm). Circles: data, Line: best fit with $K(R_{g}/R)^{b}$. (d):
$\Delta G^{(o)}_{PS-PS}$, variation of cohesion versus depth $z$,
for $R$-values shown. Circles: data, Line: best fit with modified
P\"{o}schl-Teller (MPT) function. Curves side-shifted for clarity.}

\caption{(color online) Phase images of tapping-mode Atomic Force
Microscopy (AFM) scans ($500nm \times 500nm$) of \emph{PS} films
with $R =$  50.0~nm ($\simeq$ 2$R_{g}$)(a) and  84.0~nm ($\simeq$
4$R_{g}$)(b). (c): $\Delta G^{(i)}_{PS-PS}$, variation of cohesion
versus in-plane co-ordinate $x$, for different $R$-values shown.
Circles: data, Line: best fit with MPT function. Curves have been
side-shifted for clarity. (d): Schematic of the confined system.}

\caption{(color online) Transmission spectra (absorbance versus
photon energy) in vacuum ultraviolet (vuv) for \emph{PS} films with
$R =$ 180.0~nm (a) and 50.0~nm (b). Assigned transitions presented
beside spectral bands. Circles: data, red line: convolution of
individual gaussian fits (only those for 'dimer' peaks shown in
green). Inset: Configurations of benzene 'dimers' extracted from
analysis of $^1A_{1g} \rightarrow ^1E_{1u}$. $\alpha$ = dihedral
angle.}

\end{figure}
\begin{table}
  \centering
  \caption{Parameters of the intermolecular potential}\label{rt}
  \begin{tabular}{|c|c|c|c|c|}
  \hline\hline
  Film  & \multicolumn{2}{|c|}{Peak strength }
  & \multicolumn{2}{|c|}{Range } \\
  Thickness  & \multicolumn{2}{|c|}{($V_0$)(mJ $m^{-2}$)from}
  & \multicolumn{2}{|c|}{($\Lambda$)(nm)from} \\ \cline{2-5}
  (nm)&\hspace{.1in} XR \footnotemark[1]&  AFM & \hspace{.05in} XR \hspace{.1in}&  AFM \\ \hline
  114 & 0 & 1.42& 0 & 5.6  \\
   84& 1.97 & 3.29& 5.9 & 6.5 \\
   60.4& 3.36 & 3.47 & 5.3 &5.1 \\
   58.5& 3.73 & -& 4.8 & -\\
   52 & 4.68 & - & 6.4 &-  \\
   50 & 5.36 & 6.37 & 5.3 & 5.3  \\
   48.5& 5.95 & -& 5.3 &-\\ \hline
\end{tabular}
\footnotetext[1]{X-ray Reflectivity}
\end{table}


\begin{thebibliography}{28}

\bibitem{layer} C.-J. Yu et al, Phys. Rev. Lett. {\bf 82}, 2326 (1999); M.J. Zwanenburg et al, Phys. Rev. Lett. \textbf{85}, 5154 (2000); S.E. Donnelly et al, Science \textbf{296}, 507 (2002).
\bibitem{epl96} M. K. Sanyal, J. K. Basu, A. Datta, and S. Banerjee,
Europhys. Lett., {\bf 36}, 265 (1996).
\bibitem{epl02a} O.H. Seeck et al, Europhys. Lett. \textbf{60}, 376 (2002).
\bibitem{DoiEdBk} M. Doi and S. F. Edwards, \emph{The Theory of polymer dynamics},
Clarendon Press, Oxford, 1986.
\bibitem{props} A.L. Demirel and S. Granick, J. Chem. Phys. \textbf{117}, 7745 (2001); M. Mukherjee et al, Phys. Rev. E \textbf{66}, 061801 (2002);  J. Schuster, F. Cichos and C. v. Borzcyskowski, European Polymer Journal \textbf{40}, 993 (2004).
\bibitem{rev}M. Alcoutlabi and G. B. McKenna, J. Phys.: Condens. Matter \textbf{17}, R461 (2005).
\bibitem{lan06} S. Patil, G. Matei, A. Oral and P.M. Hoffmann,
Langmuir \textbf{22}, 6484 (2006).
\bibitem{prb05} S. Chattopadhyay and A. Datta, Phys. Rev. B {\bf 72}, 155418 (2005).
\bibitem{Muthu} M. Muthukumar, Phys. Rev. Lett. {\bf 86}, 3188
(2001); C. Y. Kong and M. Muthukumar, J. Chem. Phys. {\bf 120}, 3460
(2004).
\bibitem{nl06} A. Cacciuto and E. Luijten, Nano Letters {\bf 6}, 901 (2006).
\bibitem{FlBk} S. Fl\"{u}gge, \emph{Practical Quantum Mechanics}, Springer
International Student Edition, Narosa Publishing House, New Delhi,
1979.
\bibitem{IsraBk} J. N. Israelachvili, \emph{Intermolecular and Surface Forces} (Academic
Press, New York, 1992)
\bibitem{Hoff} M. G. Cacace, E. M. Landau and J. J. Ramsden, Q. Rev.
Biophys. {\bf 30}, 241(1997).
\bibitem{Tam} J. Tamayo and R. Garcia, Appl. Phys. Lett. \textbf{71}, 2394
(1997); J. Tamayo and R. Garcia, Appl. Phys. Lett. \textbf{73}, 2926
(1998).
\bibitem{prb05a} A. Schirmeisen and H. H\"{o}lscher Phys. Rev. B {\bf 72}, 45431 (2005).
\bibitem{Langmuir04} L.-O. Heim, M. Kappl and H.-J. Butt, Langmuir \textbf{20}, 2760 (2004).
\bibitem{prl02epj03} A. Sharma and J. Mittal, Phys. Rev. Lett. {\bf 89}, 186101
(2002); C. Bollinne et al, Eur. Phys. J. E {\bf 12}, 389 (2003).
\bibitem{bear} S. Nannarone et al, AIP Conf. Proc. \textbf{705}, 450
(2004); L. Pasquali, A. De Luisa and S. Nannarone, AIP Conf. Proc.
\textbf{705},1142 (2004).
\bibitem{PopeBk} M. Pope and C. E. Swenberg, \emph{Electronic Processes In Organic
Crystals} (Oxford University Press, New York, 1982).
\bibitem{cpl04} S. Chattopadhyay and A. Datta, Chem. Phys. Lett.
\textbf{391}, 216 (2004).
\bibitem{jcp75} K.C. Janda et al, J. Chem. Phys. \textbf{63}, 1419
(1975); S. Tsuzuki et al, J. Chem. Phys. \textbf{117}, 11216 (2002).
\bibitem{macro06} V. Lupa\c{s}cu, S.J. Picken and M. W\"{u}bbenhorst, Macromolecules \textbf{39}, 5152 (2006).
\bibitem{Tg} J. A. Forrest, K. Dalnoki-Veress, J. R. Stevens, and J. R. Dutcher, Phys. Rev. Lett. \textbf{77}, 2002 (1996); C.J. Ellison, R.L. Ruszkowski, N.J. Fredin and J.M. Torkelson, Phys. Rev. Lett. \textbf{92}, 095702 (2004).


\end{thebibliography}
\end{document}